\documentclass[12pt]{article}
\usepackage{graphicx}
\usepackage{hyperref}
\usepackage{lineno}
\usepackage[OT1]{fontenc}
\usepackage{amsthm,amsmath,amssymb}
\usepackage[numbers]{natbib}
\usepackage{graphicx}
\usepackage{subfig}
\usepackage{titlesec}
\usepackage{mathrsfs}
\usepackage{multirow}
\usepackage{stackengine}
\usepackage{rotating}
\usepackage{color}
\usepackage{subfig}
\RequirePackage{enumitem} 
\RequirePackage{mathtools}
\usepackage[doublespacing]{setspace}
\usepackage{geometry}
\numberwithin{equation}{section}
\theoremstyle{plain}

\titlelabel{\thetitle.\quad}

\begin{document}

\title{A web application for the design of multi-arm clinical trials}
\author{Michael J. Grayling\textsuperscript{1*}, James M. S. Wason\textsuperscript{1,2}\\
	\small 1. Institute of Health \& Society, Newcastle University, Newcastle, UK,\\ \small 2. Hub for Trials Methodology Research, MRC Biostatistics Unit, Cambridge, UK. \\ \small *Address correspondence to M. J. Grayling, Institute of Health \& Society,\\ \small Baddiley Clark Building, Richardson Road, Newcastle upon Tyne NE2 4AX, UK.\\ \small E-mail: michael.grayling@newcastle.ac.uk.}
\date{}
\maketitle

\noindent \textbf{Abstract:} Multi-arm designs provide an effective means of evaluating several treatments within the same clinical trial. Given the large number of treatments now available for testing in many disease areas, it has been argued that their utilisation should increase. However, for any given clinical trial there are numerous possible multi-arm designs that could be used, and choosing between them can be a difficult task. This task is complicated further by a lack of available easy-to-use software for designing multi-arm trials. To aid the wider implementation of multi-arm clinical trial designs, we have developed a web application for sample size calculation when using a variety of popular multiple comparison corrections. Furthermore, the application supports sample size calculation to control several varieties of power, as well as the determination of optimised arm-wise allocation ratios. It is built using the Shiny package in the R programming language, is free to access on any device with an internet browser, and requires no programming knowledge to use. The application provides the core information required by statisticians and clinicians to review the operating characteristics of a chosen multi-arm clinical trial design. We hope that it will assist with the future utilisation of such designs in practice.\\



\section*{Background}

Drug development is becoming an increasingly expensive process, with the estimated average cost per approved new compound now standing at over \$1 bn \cite{DiMasi2016}. In no small part this is due to the high failure rate of clinical trials, in particular in phases II and III. This is particularly true in the field of oncology, where the likelihood of approval from phase I is only 5.1\% \cite{BIO2016}. Consequently, the clinical research community is constantly seeking new methods that may improve the efficiency of the drug development process.

One possible method, which has received substantial attention in recent years, is the idea to make use of multi-arm designs that compare several experimental treatments to a shared control group. Several desirable, inter-related, features of such designs have now been described. For example, the number of patients on the control treatment is typically reduced compared to conducting separate two-arm trials, and simultaneously patients are more likely to be randomized to an experimental treatment, which may help with recruitment \cite{Parmar2014,Jaki2018}. Furthermore, the overall required sample size, for the same level of power, will typically be smaller than that which would be required  if multiple two-arm trials were conducted \cite{Wason2014}. Finally, multi-arm designs offer a fair head-to-head comparison of experimental treatments in the same study \cite{Parmar2014,Jaki2018}, and the cost of assessing a treatment in a multi-arm trial is often around half of that for a separate two-arm trial \cite{Parmar2014}.

Based upon these advantages, and their experiences of utilising such designs in several oncology trials, Parmar et al. \cite{Parmar2014} make a compelling case for the need for more multi-arm designs to be used in clinical research. We are not aware of any systematic evidence on whether this has now permeated through to practice, but a simple search of PubMed Central suggests it may be the case: 859 articles have included the phrases ``multi-arm" and ``clinical trial'' since 2015, as opposed to just 273 in all years prior to this. Considering this result in combination with the findings of Baron et al. \cite{Baron2013}, who determined 17.9\% of trials published in 2009 were multi-arm, as well as the recent publication of a key guidance document on reporting results from multi-arm trials \cite{Juszczak2019}, it is clear that there is now much interest within the trials community in such designs.

However, whilst there are numerous advantages of multi-arm trials, it is important to recognise that determining a suitable design for a multi-arm clinical trial can be a substantially more complex process than for a two-arm trial. In particular, a decision must be made on how to account for the multiple comparisons that will be made. Indeed, whether the final analysis should adjust for multiplicity has been a topic of much debate within the literature. In brief, presented arguments primarily revolve around the fact that failing to account for multiplicity can substantially increase the probability of committing a type-I error. Yet, if a series of two-arm trials were conducted, no adjustment would be made to the significance level used in each trial. For brevity, we will not repeat all further arguments on this issue here, and instead refer the reader to several key discussions on multiplicity \cite{Wason2014,Rothman1990,Cook1996,Proschan2000,Bender2001,Feise2002,Hughes2005,Freidlin2008,Li2016,EMA2017,FDA2017,Howard2018}.

For the purposes of what follows in this article, the more important consideration is that when a multiple comparison correction (MCC) is to be used, one of a wide selection must actually be chosen (see, e.g., \cite{Hochberg1987,Hsu1996,Bretz2010} for an overview). MCCs vary widely in their complexity, with Bonferroni's correction often recommended because of its simplicity \cite{Juszczak2019}. However, other MCCs often perform better in terms of the operating characteristics they impart, as Bonferroni's correction is known to be conservative \cite{Proschan2000,Howard2018,Hsu1996,Sankoh2003}. A recent review found that amongst those multi-arm trials that did adjust for multiplicity, 50\% used one of the comparatively simple Bonferroni or Dunnett corrections \cite{Wason2014}. Thus, there arguably remains the potential for increased efficiency gains to be made in multi-arm trials, if more advanced MCCs can be employed.

Furthermore, regardless of whether a MCC is utilised, there are other complications that must also be addressed in multi-arm trial design, including how to power the trial, and what the allocation ratio to each experimental arm relative to the control arm will be. Indeed, power is not a simple quantity in a multi-arm trial, whilst the literature on how to choose the allocation ratios in an optimal manner is extensive (see, e.g., \cite{Atkinson2007} for an overview), and deciding whether to specify allocation ratios absolutely, or whether they can be optimised to improve trial efficiency may not be an easy decision.

These considerations imply that user-friendly software for designing multi-arm clinical trials would be a valuable tool in the trials community. It is unfortunate therefore that little software is available to assist with such studies. The principal exception to this is the MULTIARM module for East \cite{East2019}, which allows users to compare the operating characteristics of many multi-arm designs with respect to numerous important quantities. However, the cost of this package may be prohibitive to many working within academia. For this reason, we have developed a web application for multi-arm clinical trial design. We hope that the availability of this application will assist with the utilization of more advanced multi-arm designs in future clinical trials.

\section*{Implementation}

The web application is written using the Shiny package \cite{Chang2019} in the R programming language \cite{R2018}. It is available as a function in (for off-line local use), and is built using other functions from, the R package multiarm \cite{Grayling2019}. A vignette is provided for multiarm that gives great detail on its formal statistical specifications. A less technical summary is provided here.

\subsection*{Design setting}

It is assumed that outcomes $X_{ik}$ will be accrued from patients $i\in\{1,\dots,n_k\}$ on treatment arms $k\in\{0,\dots,K\}$, with arm $k=0$ corresponding to a shared control arm, and arms $k\in\{1,\dots,K\}$ to several experimental arms. Later, we provide more information on the precise types of outcome that are currently supported by the web application. The hypotheses of interest are assumed to be $H_k : \tau_k \le 0$ for $k\in\{1,\dots,K\}$. Here, $\tau_k$ corresponds to a treatment effect for experimental arm $k\in\{1,\dots,K\}$ relative to the control arm. Thus, we assume one-sided tests for superiority. Note that in the app, reference is also made to the global null hypothesis, $H_G$, which we define to be the scenario with $\tau_1=\cdots=\tau_K=0$. 

To test hypothesis $H_k$, we assume that a Wald test statistic, $z_k$, is computed
$$ z_k = \frac{\hat{\tau}_k}{\sqrt{\text{Var}(\hat{\tau}_k)}} = \hat{\tau}_kI_k^{1/2},\ \ k\in\{1,\dots,K\}. $$
In what follows, we use the notation $\boldsymbol{z}_k=(z_1,\dots,z_k)^\top\in\mathbb{R}^k$. With this, note that our app supports design in particular scenarios where $\boldsymbol{Z}_k$, the random pre-trial value of $\boldsymbol{z}_k$, has (at least asymptotically) a $k$-dimensional multivariate normal (MVN) distribution, with
\begin{align*}
\mathbb{E}(Z_l) &= \tau_lI_l^{1/2},\ l=1,\dots,k,\\
\text{Cov}(Z_l,Z_l) &= 1, \ l\in\{1,\dots,k\},\\
\text{Cov}(Z_{l_1},Z_{l_2}) &= I_{l_1}^{1/2}I_{l_2}^{1/2}\text{Cov}(\tau_{l_1},\tau_{l_2}),\ l_1 \neq l_2,\ l_1,l_2\in\{1,\dots,k\}.
\end{align*}
As is discussed further later, this includes normally distributed outcome variable scenarios and, for large sample sizes, other parametric distributions such as Bernoulli outcome data.

Ultimately, to test the hypotheses, $\boldsymbol{z}_K$ is converted to a vector of $p$-values,
$\boldsymbol{p}=(p_1,\dots,p_K)^\top\in[0,1]^K$, via $p_k = 1 - \Phi_1(z_k,0,1)$, for $k\in\{1,\dots,K\}$. Here, $\Phi_n\{(a_1,\dots,a_n)^\top,\boldsymbol{\lambda},\Sigma\}$ is the cumulative distribution function of an $n$-dimensional MVN distribution, with mean $\boldsymbol{\lambda}$ and covariance matrix $\Sigma$. Precisely
$$ \Phi_n\{(a_1,\dots,a_n)^\top,\boldsymbol{\lambda},\Sigma\} = \int_{-\infty}^{a_1}\dots\int_{-\infty}^{a_n}\phi_n\{\boldsymbol{x},\boldsymbol{\lambda},\Sigma\}\mathrm{d}x_n\dots\mathrm{d}x_1, $$
where $\phi_n\{\boldsymbol{x},\boldsymbol{\lambda},\Sigma\}$ is the probability density function of an $n$-dimensional MVN distribution with mean $\boldsymbol{\lambda}$ and covariance matrix $\Sigma$, evaluated at vector $\boldsymbol{x}=(x_1,\dots,x_n)^\top$.

Then, which null hypotheses are rejected is determined by comparing the $p_k$ to a set of significance thresholds specified based on a chosen MCC, in combination with a nominated significance level $\alpha\in(0,1)$. Before we describe the currently supported MCCs however, we will first describe the operating characteristics that are currently evaluated by the app.

\subsection*{Operating characteristics}

Our app returns a wide selection of statistical operating characteristics that may be of interest when choosing a multi-arm trial design. Specifically, it can compute the following quantities for any nominated multi-arm design and true set of treatment effects

\begin{itemize}
\item The conjunctive power ($P_\text{con}$): The probability that all of the null hypotheses are rejected, irrespective of whether they are true or false.
\item The disjunctive power ($P_\text{dis}$): The probability that at least one of the null hypotheses is rejected, irrespective of whether they are true or false.
\item The marginal power for arm $k\in\{1,\dots,K\}$ ($P_k$): The probability that $H_k$ is rejected, irrespective of whether it is true or false.
\item The per-hypothesis error-rate ($PHER$): The expected value of the number of type-I errors divided by the number of hypotheses.
\item The $a$-generalised type-I familywise error-rate ($FWER_{Ia}$): The probability that at least $a\in\{1,\dots,K\}$ type-I errors are made. Note that $FWER_{I1}$ is the conventional familywise error-rate ($FWER$); the probability of making at least one type-I error.
\item The $a$-generalised type-II familywise error-rate ($FWER_{IIa}$): The probability that at least $a\in\{1,\dots,K\}$ type-II errors are made.
\item The false discovery rate ($FDR$): The expected proportion of type-I errors amongst the rejected hypotheses.
\item The false non-discovery rate ($FNDR$): The expected proportion of type-II errors amongst the hypotheses that are not rejected.
\item The positive false discovery rate ($pFDR$): The rate that rejections are type-I errors.
\item The sensitivity ($Sensitivity$): The expected proportion of the number of correct rejections of the hypotheses to the number of false null hypotheses.
\item The specificity ($Specificity$): The expected proportion of the number of correctly not rejected hypotheses to the number of true null
hypotheses.
\end{itemize}

\subsection*{Multiple comparison corrections}

\subsubsection*{Per-hypothesis error-rate control}

The most simple method for selecting the significance thresholds
against which to compare the $p_k$, is to compare each to the chosen
significance level $\alpha$. That is, to reject $H_k$ for $k\in\{1,\dots,K\}$ if
$p_k \le \alpha$. This controls the $PHER$ to $\alpha$.

A potential problem with this, however, can be that the statistical operating characteristics of the resulting design may not be desirable (e.g., in terms of $FWER_{I1}$). As discussed earlier, it is for this reason that we may wish to make use of a MCC. Currently, the web application supports the use of a variety of such MCCs, which aim to control either (a) the conventional familywise error-rate, $FWER_{I1}$ (with these techniques sub-divided into single-step, step-down, and step-up corrections) or (b) the $FDR$.

\subsubsection*{Single-step familywise error-rate control}

These MCCs test each of the $H_k$ against a common significance level,
$\gamma\in(0,1)$ say, rejecting $H_k$ if $p_k \le \gamma$. The currently supported single-step corrections are

\begin{itemize}
\item Bonferroni's correction: This sets $\gamma = \alpha/K$ \cite{Bonferroni1936}.
\item Sidak's correction: This sets $\gamma = 1 - (1 - \alpha)^{1/K}$ \cite{Sidak1967}.
\item Dunnett's correction: This sets $\gamma = 1 - \Phi_1\{z_D,0,1\}$, where $z_\text{D}$ is the solution of the following equation
$$ \alpha = 1 - \Phi_K\{(z_D,\dots,z_D)^\top,\boldsymbol{0}_K,\text{Cov}(\boldsymbol{Z}_K)\}, $$
with $\boldsymbol{0}_n=(0,\dots,0)^\top\in\mathbb{R}^n$ an $n$-dimensional vector of zeroes \cite{Dunnett1955}.
\end{itemize}

Note that each of the above specify a $\gamma$ such that the maximum probability of incorrectly rejecting at least one of the null hypotheses $H_k$, $k\in\{1,\dots,K\}$, over all possible values of $\boldsymbol{\tau}\in\mathbb{R}^K$ is at most $\alpha$. This is referred to as strong control of $FWER_{I1}$.

\subsubsection*{Step-down familywise error-rate control}

Step-down MCCs work by ranking the $p$-values from smallest to largest. We will refer to these ranked $p$-values by $p_{(1)},\dots,p_{(K)}$, with associated hypotheses $H_{(1)},\dots,H_{(K)}$. The $p_{(k)}$ are compared to a vector of significance levels $\boldsymbol{\gamma}=(\gamma_1,\dots,\gamma_K)\in(0,1)^K$.
Precisely, the maximal index $k$ such that $p_{(k)}>\gamma_k$ is identified, and then $H_{(1)},\dots,H_{(k-1)}$ are rejected and $H_{(k)},\dots,H_{(K)}$ are not rejected. If $k=1$ then we do not reject any of the null hypotheses, and if no such $k$ exists then we reject all of the null hypotheses. The currently supported step-down corrections are

\begin{itemize}
\item Holm-Bonferroni correction: This sets $\gamma_k=\alpha/(K+1-k)$ \cite{Holm1979}.
\item Holm-Sidak correction: This sets $\gamma_k = 1 - (1 - \alpha)^{K+1-k}$.
\item Step-down Dunnett correction: This can only currently be used when the
$\text{Cov}(Z_{k_1},Z_{k_2})$ are equal for all
$k_1\neq k_2,\ k_1,k_2\in\{1,\dots,K\}$. In this case, it sets
$\gamma_k = 1 - \Phi_1\{z_{Dk},0,1\}$, where $z_{Dk}$ is the solution to
$$ \alpha = 1 - \Phi_{K+1-k}\{(z_{Dk},\dots,z_{Dk})^\top,\boldsymbol{0}_{K+1-k},\text{Cov}(\boldsymbol{Z}_{K+1-k})\}. $$
\end{itemize}

Note that both of the above methods provide strong control of $FWER_{I1}$.

\subsubsection*{Step-up familywise error-rate control}

Step-up MCCs also work by ranking the $p$-values from smallest to largest, and similarly utilise a vector of significance levels $\boldsymbol{\gamma}$. However, here, the largest $k$ such that $p_{(k)} \le \gamma_k$ is identified. Then, the hypotheses $H_{(1)},\dots,H_{(k)}$ are rejected, and $H_{(k+1)},\dots,H_{(K)}$ are not rejected. Currently, one such correction is supported: Hochberg's correction \cite{Hochberg1988}, which sets $\gamma_k=\alpha/(K+1-k)$. This method also provides strong control of $FWER_{I1}$.

\subsubsection*{False discovery rate control}

It may be of interest to instead control the $FDR$, which can offer a compromise between strict $FWER_{I1}$ control and $PHER$ control, especially when we expect a large proportion of the experimental treatments to be effective. Currently, two methods that will control the $FDR$ to at most $\alpha$ over all possible $\boldsymbol{\tau}\in\mathbb{R}^K$ are supported. They function in the same way as the step-up corrections discussed above, with

\begin{itemize}
\item Benjamini-Hochberg correction: This sets $\gamma_k=k\alpha/K$ \cite{Benjamini1995}.
\item Benjamini-Yekutieli correction: This sets \cite{Benjamini2001}:
$$\gamma_k=\frac{k\alpha}{K\left(1 + \frac{1}{2} + \dots + \frac{1}{K}\right)}.$$
\end{itemize}

\subsection*{Sample size determination}

The sample size required by a design to control several types of power to a specified level $1-\beta$, under certain specific scenarios, can be computed. Precisely, following for example \cite{Wason2016}, values for `interesting' and `uninteresting' treatment effects, $\delta_1\in\mathbb{R}^+$ and $\delta_0\in(-\infty,\delta_1)$ respectively, are specified and the following definitions are made

\begin{itemize}
\item The global alternative hypothesis, $H_A$, is given by $\tau_1=\cdots=\tau_K=\delta_1$.
\item The least favourable configuration for experimental arm $k\in\{1,\dots,K\}$, $LFC_k$, is given by $\tau_k=\delta_1,\ \tau_1=\cdots=\tau_{k-1}=\tau_{k+1}=\cdots=\tau_K=\delta_0$.
\end{itemize}

Then, the following types of power can be controlled to level $1-\beta$ by design's determined using the app

\begin{itemize}
\item The conjunctive power under $H_A$.
\item The disjunctive power under $H_A$.
\item The minimum marginal power under the respective $LFC_k$.
\end{itemize}

\subsection*{Allocation ratios}

One of the primary goals of the app is to aid the choice of values for $n_0,\dots,n_K$. The app specifically supports the determination of values for these parameters by searching for a suitable $n_0$ via a one-dimensional root solving algorithm, and then sets $n_k=r_kn_0$, $r_k\in(0,\infty)$, for $k\in\{1,\dots,K\}$. Here, $r_k$ is the allocation ratio for experimental arm $k$ relative to the control arm.

For this reason, the app also allows the allocation ratios to be specified in a variety of ways: they can be defined explicitly, or alternatively can be determined in an optimal manner. For this optimality problem, many possible optimality criteria have been defined, each with their own merits. Therefore, we refer the reader to Atkinson (2007) \cite{Atkinson2007} for further details of optimal allocation in multi-arm designs. Instead, we simply note that in the web application, the allocation ratios can currently be determined for three such criteria

\begin{itemize}
\item $A$-optimality: Minimizes the trace of the inverse of the information matrix of the design. This results in the minimization of the average variance of the treatment effect estimates.
\item $D$-optimality: Maximizes the determinant of the information matrix of the design. This results in the minimization of the volume of the confidence ellipsoid for the treatment effect estimates.
\item $E$-optimality: Maximizes the minimum eigenvalue of the information matrix. This results in the minimization of the maximum variance of the treatment effect estimates.
\end{itemize}

The optimal allocation ratios are identified in the app using available closed-form solutions were possible (see \cite{Sverdlov2013} for a summary of these), otherwise non-linear programming is employed.

\subsection*{Other design specifications}

Finally, the web application also supports the following options

\begin{itemize}
\item Plot production: Plots can be produced of (a) all of the operating characteristics quantities listed earlier when $\tau_1=\dots=\tau_K=\theta$, as well as (b) the $P_k$ when $\tau_k=\theta$ and $\tau_l=\theta-(\delta_1-\delta_0)$ for $l\neq k$. If these are selected for rendering, the quality of the plots, in terms of the number of values of $\theta$ used for line-graph production, can also be controlled.
\item Require $n_k\in\mathbb{N}$ for $k\in\{0,\dots,K\}$: By default, the sample size determined for each arm will only be required to be a positive number. In practice, such values need to be integers. This can thus be enforced if desired, with the integer $n_k$ specified by rounding up their determined continuous values.
\end{itemize}

\subsection*{Supported outcome variables}

\subsubsection*{Normally distributed outcome variables}

Currently, the app supports multi-arm trial design for scenarios in which
the outcome variables are assumed to be either normally or Bernoulli
distributed.

Precisely, for the normal case, it assumes that
$X_{ik}\sim N(\mu_k,\sigma_k^2)$, and that $\sigma_k^2$ is known for
$k\in\{0,\dots,K\}$. Then, for each $k\in\{1,\dots,K\}$
\begin{align*}
\tau_k &= \mu_k-\mu_0,\\
\hat{\tau}_k &= \frac{1}{n_k}\sum_{l=1}^{n_k}x_{ik} - \frac{1}{n_0}\sum_{l=1}^{n_0}x_{i0},\\
I_{k} &= \frac{1}{\frac{\sigma_0^2}{n_0} +
\frac{\sigma_k^2}{n_k}},
\end{align*}
where $x_{ik}$ is the realised value of $X_{ik}$.

Note that in this case, $\boldsymbol{Z}_K$ has a MVN distribution, and thus the operating characteristics can be computed exactly and efficiently using MVN integration \cite{Bretz2019}. Furthermore, the distribution of $\boldsymbol{Z}_K$ does not depend upon the values of the $\mu_k$, $k\in\{0,\dots,K\}$. Consequently, these parameters play no part in the inputs or outputs of the app.

\subsubsection*{Bernoulli distributed outcome variables}

In this case, $X_{ik}\sim Bern(\pi_k)$ for response rates $\pi_k$, and for each $k\in\{1,\dots,K\}$
\begin{align*}
\tau_k &= \pi_k-\pi_0,\\
\hat{\tau}_k &= \frac{1}{n_k}\sum_{l=1}^{n_k}x_{ik} - \frac{1}{n_0}\sum_{l=1}^{n_0}x_{i0},\\
I_{k} &= \frac{1}{\frac{\pi_0(1 - \pi_0)}{n_0} +
\frac{\pi_k(1 - \pi_k)}{n_k}}.
\end{align*}
Thus, a problem for design determination becomes that the $I_{k}$ are
dependent on the unknown response rates. In practice, this is handled at the analysis stage of a trial by setting
$$ I_k = \frac{1}{\frac{\hat{\pi}_0(1 - \hat{\pi}_0)}{n_0} +
\frac{\hat{\pi}_k(1 - \hat{\pi}_k)}{n_k}},$$
for $\hat{\pi}_k = \sum_{l=1}^{n_k}x_{ik}/n_k$, $k\in\{0,\dots,K\}$. This is the assumption made where required in by the app. With this, $\boldsymbol{Z}_K$ is only asymptotically MVN. Thus, in general it would be important to validate operating characteristics evaluated using MVN integration via simulation.

In addition, note that the above problem also means that the operating
characterstics under $H_G$, $H_A$, and the $LFC_k$ are not unique without further restriction. Thus, to achieve uniqueness, the app requires a value be
specified for $\pi_0$ for use in the definition of these scenarios. Moreover, for this reason, the inputs and outputs of functions supporting Bernoulli
outcomes make no reference to the $\tau_k$, and work instead directly in terms
of the $\pi_k$. Finally, note that this problem also means that to determine $A$-, $D$-, or $E$-optimised allocation ratios, a specific set of values for the $\pi_k$ must be assumed.

In this case, we should also ensure that $\delta_1\in(0,1)$ and $\delta_0\in(-\pi_0,\delta_1)$, for the assumed value of $\pi_0$, since $\pi_k\in[0,1]$ for $k\in\{1,\dots,K\}$.

\section*{Results}

\subsection*{Support}

The web application is freely available from \href{https://mjgrayling.shinyapps.io/multiarm/}{https://mjgrayling.shinyapps.io/multiarm/}. The R code for the application can also be downloaded from \href{https://github.com/mjg211/multiarm}{https://github.com/mjg211/multiarm}. Furthermore, as noted earlier, the app is built in to the package multiarm \cite{Grayling2019}, as the function \texttt{gui()}, for ease-of-use without internet access. The application has a simple interface, and has the capability to

\begin{itemize}
\item Determine the sample required in each arm in a specified multi-arm clinical trial design scenario;
\item Summarise and plot the operating characteristics of the identified design;
\item Produce a report describing the chosen design scenario, the identified design, and a summary of its operating characteristics.
\end{itemize}

\subsection*{Inputs}

The outputs (i.e., the identified design and its operating characteristics) are determined based upon the following set of user specified inputs (Figure 1)

\begin{enumerate}
\item The number of experimental treatment arms, $K$.
\item The chosen multiple comparison correction (e.g., Dunnett's correction).
\item The significance level, $\alpha$.
\item The type of power to control (e.g., the conjunctive power under $H_A$).
\item The desired power, $1-\beta$.
\item For Bernoulli distributed data, the control arm response rate $\pi_0$.
\item The interesting treatment effect, $\delta_1$.
\item The uninteresting treatment effect, $\delta_0$.
\item For normally distributed data, the standard deviations, $\sigma_0,\dots,\sigma_K$. These are allocated by first selecting the type of standard deviations (e.g., that they are assumed to be equal across all arms), and then the actual values for the parameters.
\item The allocation ratios (e.g., $A$-optimal).
\item For Bernoulli distributed data, when searching for optimal allocation ratios, the response rates to assume in the search.
\item Whether the sample size in each arm should be required to be an integer;
\item Whether plots should be produced, and if so the plot quality.
\end{enumerate}

Note that a \textit{Reset inputs} button is provided to simplify returning the inputs to their default values. Once the inputs have been specified as desired, the outputs can be generated by clicking the \textit{Update outputs} button.

\subsection*{Example}

Here, we demonstrate specification of the input parameters (Figure 1), and then subsequent output generation (Figures 2-4), for parameters motivated by a three-arm phase II randomized controlled trial of treatments for myelodysplastic syndrome patients, described in \cite{Jacob2016}. This trial compared, via a binary primary outcome, two experimental treatments with conventional azacitidine treatment. The trial was designed with $\alpha=0.15$, $\beta=0.2$, $\delta_1=0.15$, and $\pi_0=0.3$. For simplicity, we assume that the familiar Dunnett correction will be used, that $\delta_0=0$, and that allocation will be equal across the arms ($r_1=\cdots=r_K=1$). Finally, we assume it is the minimum marginal power that should be controlled.

Each input widget in Figure 1 can be seen to have been allocated accordingly based on the description above, whilst we have additionally elected to produce plots (of medium quality), and to not require the arm-wise sample sizes to be integers. Note that in Figure 1 we can see that the input widgets are supported by help boxes that can be opened by clicking on the small question marks beside them.

Figure 2 then depicts the output to the \textit{Design summary} box once the user clicks on \textit{Update outputs}. Specifically, a summary of the chosen inputs and the identified design is rendered. Furthermore, in Figure 3 we can see the tables that provide the various statistical quantities under $H_G$, $H_A$, the $LFC_k$, as well as the various treatment effect scenarios that are considered for plot production.

Finally, in Figure 4 the plots discussed earlier are shown. Observe that horizontal and vertical lines are added at the values $\alpha$, $1-\beta$, $\delta_1$, and $\delta_0$ respectively. Note that these plots are outputted in a manner to allow the user to zoom in on a particular sub-component if desired.

In all, Figures 2-4 provide a set of outputs with a variety of features that should be anticipated given the chosen input parameters. Firstly, the specification that the allocation to all arms should be equal means that $n_0=\dots=n_K$. In addition, $FWER_{I1}$ is equal to 0.15 under $H_G$, and the minimum marginal power is 0.799, as is approximately desired. Moreover, the specification that $r_1=\dots=r_K$ means that $P_\text{con}$ and $P_\text{dis}$ are equal for each of the $LFC_k$, and $P_1=P_2$.

Finally, as noted above, and as can be seen in Figure 1, a \textit{Generate report} button is provided that can produce a copy of the outputs in either PDF (.pdf), HTML (.html), or Word (.docx) format. The user can also nominate a name for this file in the \textit{Report filename} input widget. This allows a record of designs to be stored, presented, and compared to other designs if required.

\section*{Conclusions}

A possible barrier to previous calls for increased use of multi-arm clinical trial designs is a lack of available easy-to-access user-friendly software that facilitates associated sample size calculations. For this reason, we have created an online web application that supports multi-arm trial design determination for a wide selection of possible input parameters. Its use requires no knowledge of statistical programming languages and is facilitated via a simple user interface. Furthermore, we have made the application available on the internet, so that it is readily accessible, and have also made it freely available for download for remote use without an internet connection. Like similar applications that have been released recently for phase I clinical trial design \cite{Wheeler2016,Wages2018}, we hope that the availability of this application will assist with the design of future multi-arm studies.

Before we conclude, however, it is important to acknowledge the limitations of our work. Firstly, MVN integration is utilised in all instances to determine the statistical operating characteristics of potential multi-arm designs. This makes the execution time for returning outputs with many possible input parameters fast. However, there is an unavoidable complexity in certain multi-arm designs, which may make execution time long. This is particularly true of scenarios with $K \ge 5$. It can also be true of designs that utilise the more complex step-wise MCCs. It is for this reason that the application places an upper cap in the inputs of $K=5$, and also returns a warning in scenarios for which a lengthy execution time would be anticipated. Nonetheless, users may have to wait several minutes in certain situations to identify their desired design.

Furthermore, it is crucial that all software for clinical trial design be validated. This is challenging for multi-arm designs because of the aforementioned limited freely available software for designing such studies. We compared the output of our application to that of PASS \cite{PASS2019}, a validated software package, for a variety of supported input parameters, but output for many possible inputs remained difficult to corroborate because of a lack of equivalent available functionality. For this reason, we have carefully followed recommended good-programming practices and perform all statistical calculations within the application by calling functions from the R package multiarm, in which the code has been modularised \cite{Grayling2019}. Furthermore, in this package we have created a function that simulates multi-arm clinical trials using a given design. This allows us to perform an additional check on our analytical computations. Specifically, we generated 1000 random combinations of possible input parameters for trials assuming normally distributed outcomes, thus covering an extremely wide range of supported design scenarios. The analytical operating characteristics returned by the web application in the \textit{Operating characteristics summary} boxes for $H_G$, $H_A$, and the $LFC_k$ were then compared to those based on trial simulation, using 100,000 replicate simulations in each of the 1000 designs. Across all considered scenarios, the maximum absolute difference between the analytical and simulated operating characteristics was just $5\times10^{-3}$, which is within what would be anticipated due to simulation error. Consequently, it does appear that our command is functioning as desired. Code to replicate this work is available upon request from the corresponding author.

Finally, we note one primary possible avenue for future development of the web application: numerous papers have now provided designs for adaptive multi-arm trials (e.g., \cite{Magirr2012,Wason2017}), and software for their determination in certain settings \cite{Barthel2009,Jaki2019}. Given the evidential increased interest in such designs \cite{Dimairo2018}, allowing for their determination would be a valuable extension to our application.


\section*{Funding}

This work was supported by the Medical Research Council [grant number MC\_UU\_00002/6 to JMSW]. The funding body did not have any role in the design of this study, collection, analysis, and interpretation of data, nor in the writing of the manuscript.


\bibliographystyle{bmc-mathphys} 
\bibliography{multiarm}      




\begin{figure}[h!]
\centering
\includegraphics[width = 7cm]{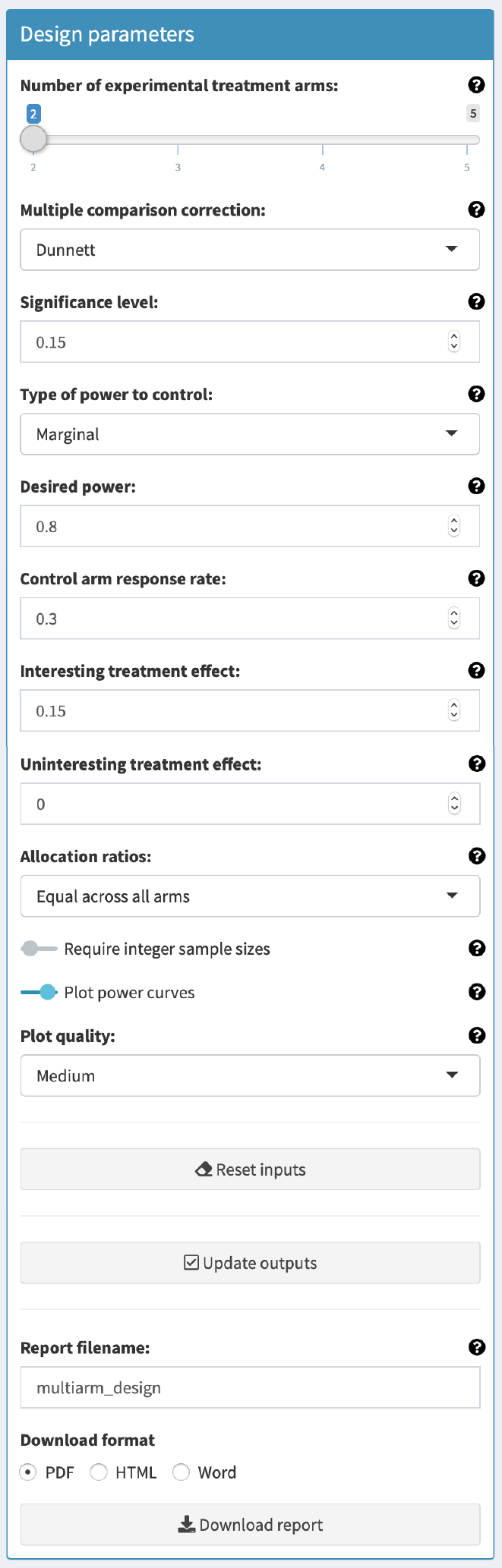}
\caption{Design parameters box. The box in which input parameters are specified is shown. The specific values that can be seen are those that correspond to the trial described in \cite{Jacob2016}.}
\end{figure}

\begin{figure}[h!]
\centering
\includegraphics[width = 12cm]{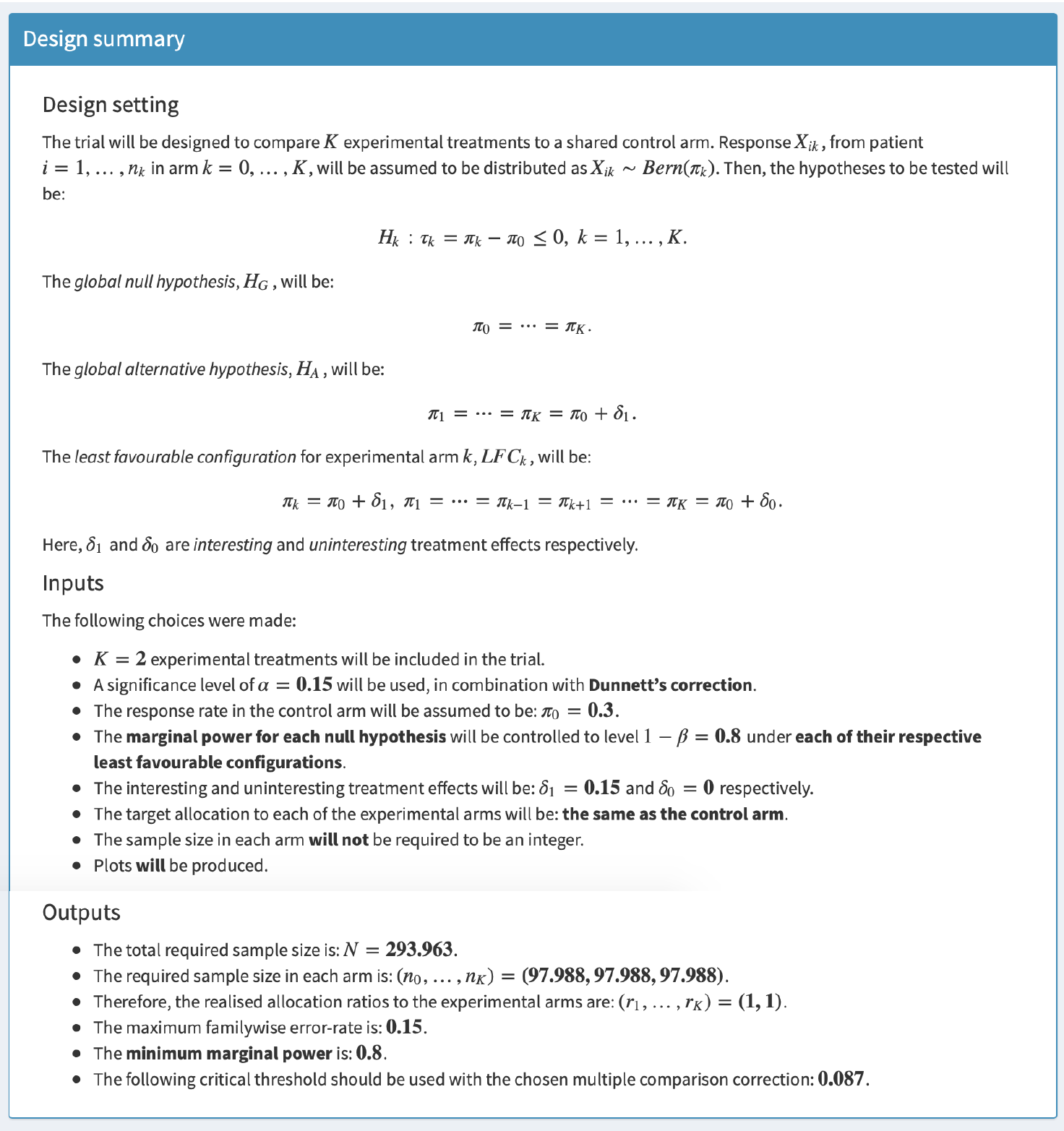}
\caption{Design summary box. The box in which a summary of the input parameters and of the identified design is rendered is shown. The specific output that can be seen corresponds to the inputs from Figure 1.}
\end{figure}

\begin{figure}[h!]
\centering
\includegraphics[width = 12cm]{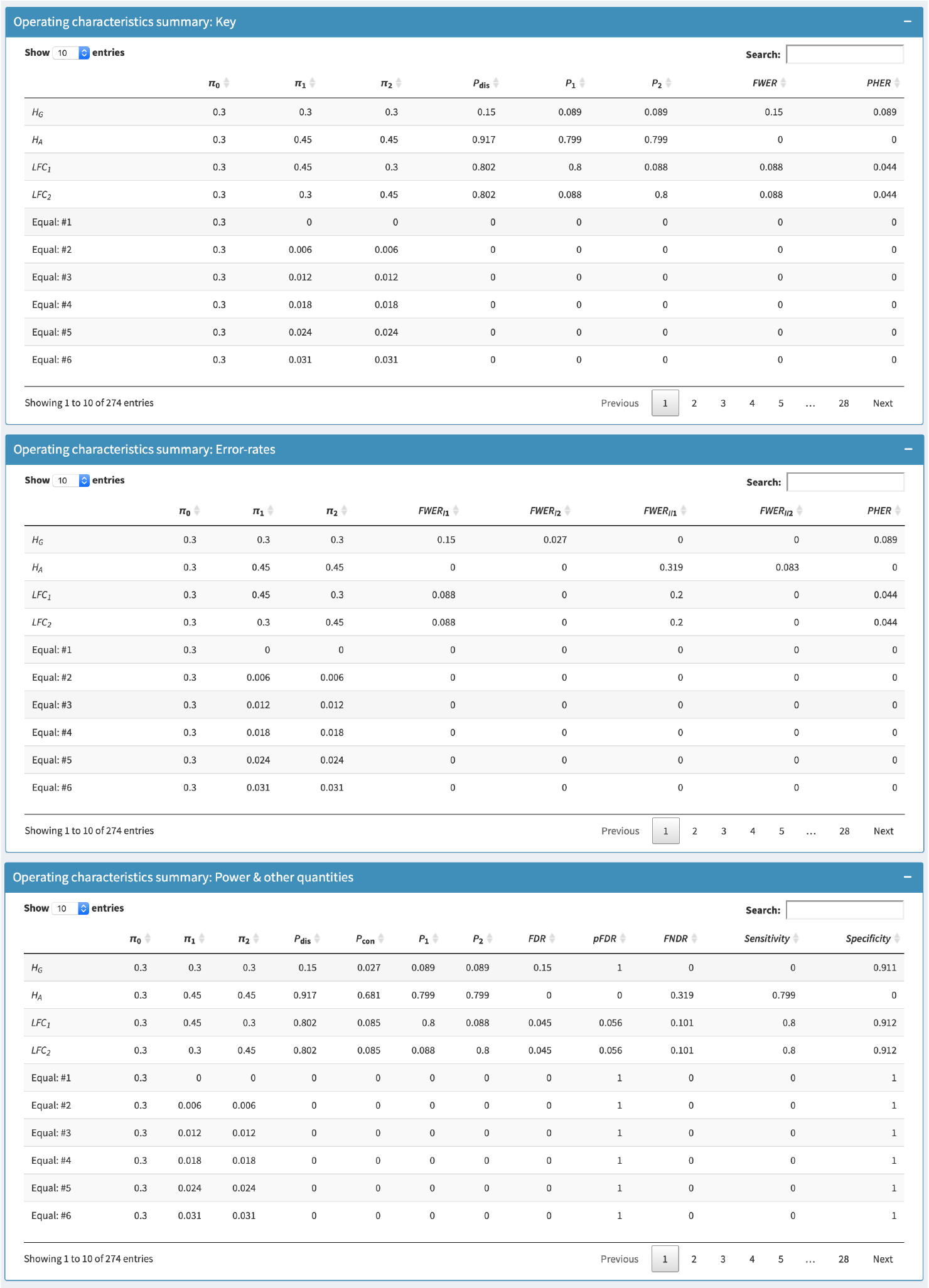}
\caption{Operating characteristics summary. The boxes in which a summary of the identified designs operating characteristics is produced is shown. The specific output that can be seen corresponds to the inputs from Figure 1.}
\end{figure}

\begin{figure}[h!]
\centering
\includegraphics[width = 12cm]{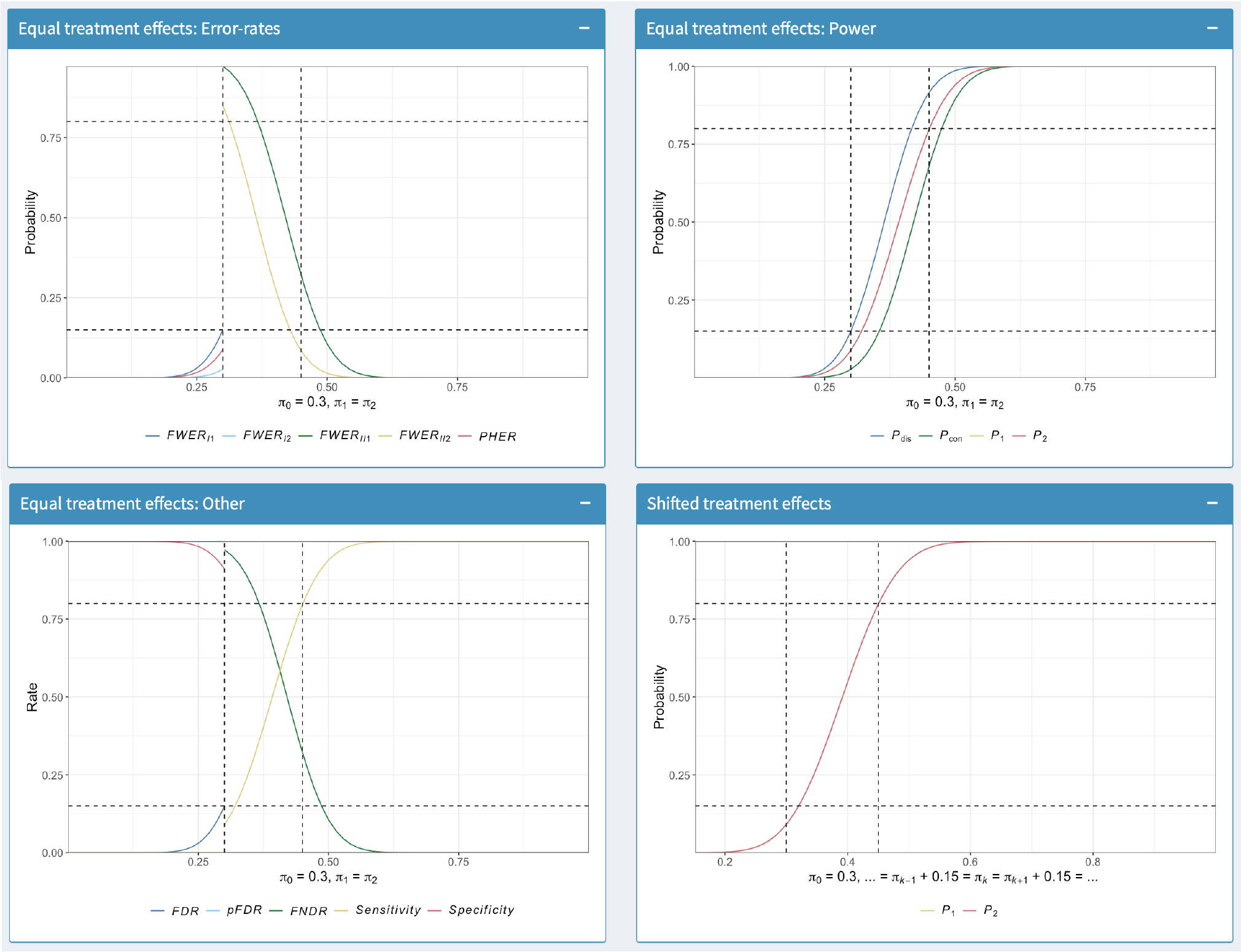}
\caption{Operating characteristics plots. The boxes in which plots of the identified designs operating characteristics are produced is shown. The specific output that can be seen corresponds to the inputs from Figure 1.}
\end{figure}




\end{document}